\documentclass[prd,aps,twocolumn,a4paper,floatfix]{revtex4}

\def\br{\nonumber \\ &&}
\def\scri{{\mathscr I}} 
\usepackage{graphicx,psfrag}
\usepackage{mathrsfs}

\psfrag{i+R}{$i^+_R$}
\psfrag{i0R}{$i^0_R$}
\psfrag{i-R}{$i^-_R$}
\psfrag{i+L}{$i^+_L$}
\psfrag{i0L}{$i^0_L$}
\psfrag{i-L}{$i^-_L$}
\psfrag{scri+R}{$\scri^+_R$} 
\psfrag{scri-R}{$\scri^-_R$}
\psfrag{scri+L}{$\scri^+_L$}
\psfrag{scri-L}{$\scri^-_L$}
\psfrag{L}{L}
\psfrag{R}{R}
\psfrag{F}{F}
\psfrag{P}{P}

\begin{document}


\title{Comments on Bona-Mass\'o type slicing conditions in long-term black hole
evolutions}

\author{David Garfinkle}

\affiliation{Department of Physics, Oakland University, Rochester, MI
  48309, USA}

\author{Carsten Gundlach}

\author{David Hilditch}

\affiliation{School of Mathematics, University of Southampton,
Southampton, SO17 1BJ, UK}


\begin{abstract}

We review in generality why time-independent endstates can be reached
in black hole and collapse simulations, with and without excision.  We
characterise the Killing states of the Bona-Mass\'o slicing
condition with time derivative along the normals to the slice (``BMn'') as
solutions of a mixed elliptic/hyperbolic differential equation on the
slice. We show numerically that these steady states can be reached as
end states from typical initial data with excision but can be 
reached with the
puncture method only if the puncture is not numerically well
resolved. During the evolution, BMn
slicings often form gauge shocks. It may be that these are not seen
in current 3D simulations only through lack of resolution, although we
expect that they can be avoided with some care. Finally we point out
that excision with BMn as currently implemented is ill-posed and
therefore not expected to converge; this can be cured. In
technical appendixes, we derive the equations of pure gauge systems on
a fixed spacetime, and bring the BSSN/NOR equations into 3-dimensional
tensor form suitable for multiple coordinate patches or spherical
polar coordinates.

\end{abstract}

\maketitle

\tableofcontents


\section{Introduction}


In 2005, parallel breakthroughs in the long-term stable simulation of
binary black holes were made using two rather different approaches:
Pretorius \cite{Pretorius} used modified harmonic slicing, black holes
created in collapse, and singularity excision, while the Brownsville
\cite{Brownsville} and Goddard \cite{Goddard} groups used eternal
black holes, and singularity-avoiding (1+log) slicing. Nevertheless,
a key ingredient in both these successes is their gauge choice.

Generalising and extending recent work by Hannam {\it et al.}
\cite{Hannametal}, we further investigate the application to black
hole and collapse spacetimes of the Bona-Mass\'o slicing condition with
time derivative along the slice normals (``BMn'').  This family
includes both the ``1+log'' slicing used in \cite{Brownsville,Goddard}
and the harmonic slicing, a variant of which is use in
\cite{Pretorius}.

A desirable property for a gauge choices is that the metric becomes
time-independent to the extent that the spacetime becomes stationary
\cite{GarfinkleGundlach}. In Sec.~\ref{section:general} we explain 
carefully why
this is possible both when the black holes are excised and when a
singularity-avoiding slicing is used. We characterise Killing
coordinates geometrically in Sec.~\ref{section:Killingcoordinates}. To
fix notation, we review various lapse conditions of Bona-Mass\'o type
in Sec.~\ref{section:gaugeconditions}. In
Sec.~\ref{section:Killingcompatible} we classify Killing slicings
compatible with BMn slicing, and in particular the spherical Killing
slicings of Schwarzschild spacetime. In
Sec.~\ref{section:vacuumevolutions} we investigate numerically if any
such Killing states are in fact attractors in evolutions of the
Schwarzschild spacetime. We consider both slices with wormhole topology and
slices which end at an excision boundary inside the black hole. In
Sec.~\ref{section:scalarfieldcollapse}, we present spherically
symmetric simulations of scalar field collapse as a toy model for
black holes formed in collapse. From our mathematical and numerical
observations in these sections, we suggest improvements to current
methods for binary black hole evolutions in
Sec.~\ref{section:conclusions}.


\section{Numerical evolution of black hole spacetimes} 
\label{section:general}


\subsection{Eternal and collapse black holes}

\begin{figure}
\begin{center}
\includegraphics[width=8cm]{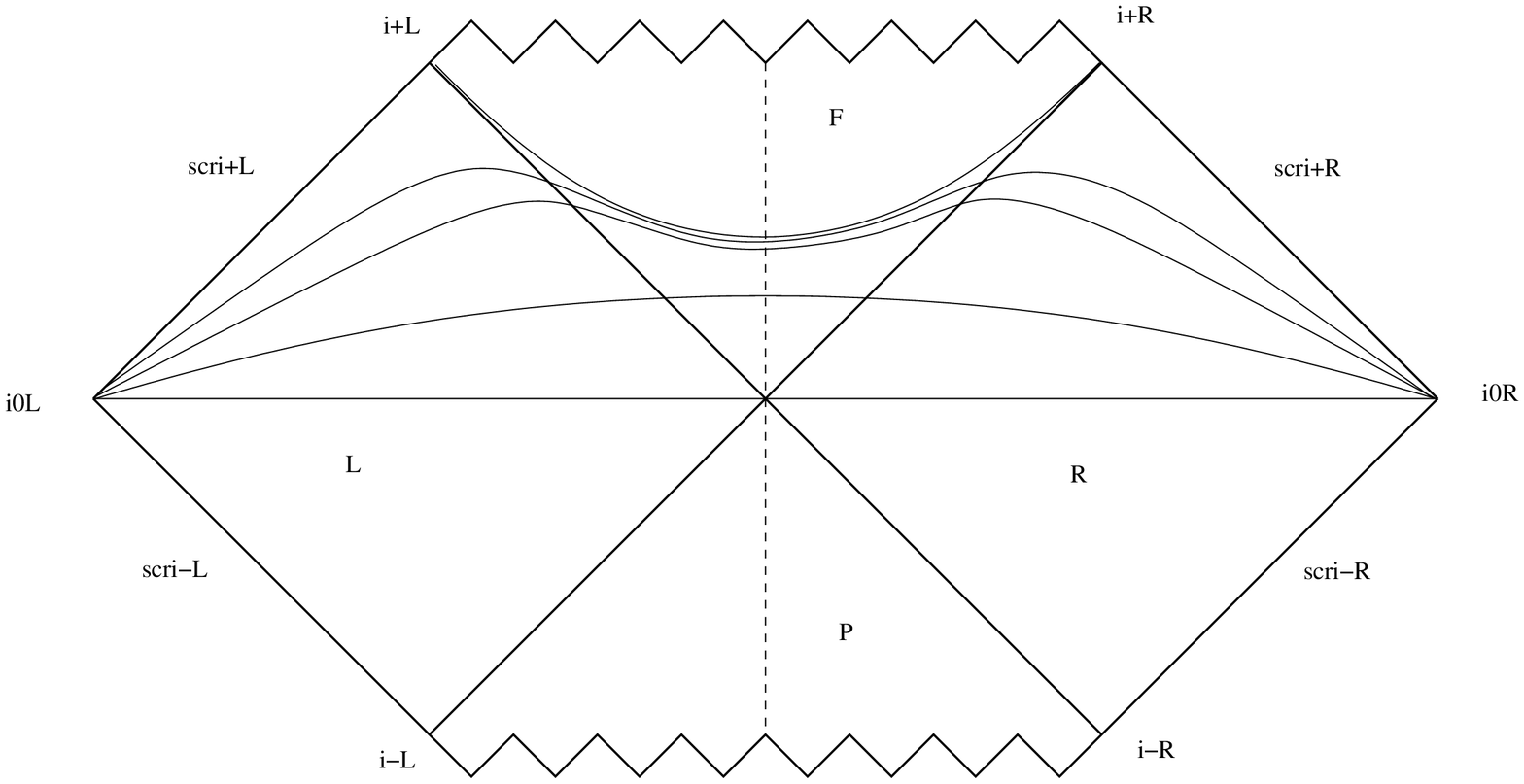}
\end{center}
\caption{ Spacetime diagram of the Schwarzschild spacetime, with the
angular coordinates suppressed. The
horizontal line from $i^0_L$ to $i^0_R$ is the time-symmetric wormhole
slice typically used as initial data in puncture evolutions of a
Schwarzschild black holes.  The curved lines schematically represent
the slicing generated from these initial data by BMn lapse with
$\alpha=1$ initially. They approach the slice $R=R_0$, which links
$i^+_L$ to $i^+_R$. The vertical dashed line represents the symmetry
boundary which can replace the left-right reflection symmetry of this
slicing. As the slices approach $R=R_0$, the approximately cylindrical
wormhole grows longer linearly with time.
\label{fig1}
}
\end{figure}

Black holes in the real world have formed in collapse, but eternal
black holes are often used in numerical relativity because they differ
from collapse black holes only in the interior, and this cannot affect
physics outside. Here we concentrate on non-rotating, uncharged black
holes, which are described by the Kruskal extension of the
Schwarzschild spacetime.  A bifurcate Killing horizon divides this
spacetime into past (P), future (F), ``left'' (L) and ``right'' (R)
regions. The future and past timelike ($i^+$ and $i^-$) and null
($\scri^+$ and $\scri^-$) infinities and the spacelike infinity $i^0$
all exist in left ($L$) and right ($R$) copies. Slices extending from
$i^0_L$ to $i^0_R$ have wormhole geometry, see Fig.~\ref{fig1}.
Binary (or multiple) black hole initial data can be represented by a
wormhole leading to a separate copy of $i^0_L$ for each black hole. In
the ``puncture'' method \cite{BrandtBruegmann}, each $i^0_L$ is then
represented in coordinates by a point where the conformal factor
diverges.

By contrast, black holes formed from regular data through collapse
have trivial spatial topology, similar to the Schwarzschild spacetime but with
part of R and F, and all of P and L, covered up by the collapsing
star \cite{OppenheimerSnyder} -- see Fig.~\ref{fig4}.

\begin{figure}
\begin{center}
\includegraphics[width=8cm]{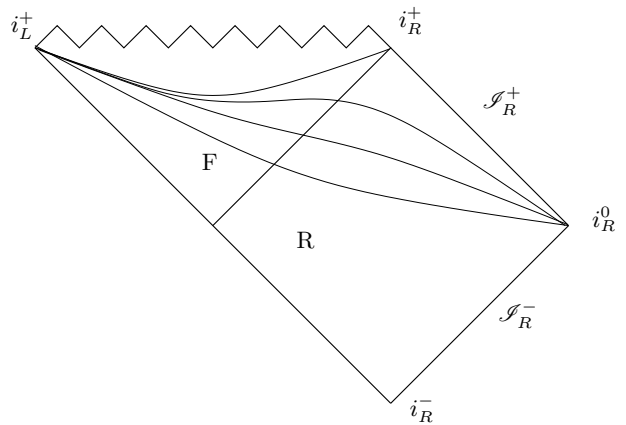}
\end{center}
\caption{
\label{fig2}
The same spacetime diagram, schematically showing the
unique regular spherical Killing slicing that is compatible with BMn
slicing (for a given $\mu_L(\alpha)$. All slices are isometric to one
another, and connect $i^+_L$ with $i^0_R$. The again asymptote to the
slice $R=R_0$.}
\end{figure}


\begin{figure}
\begin{center}
\includegraphics[width=8cm]{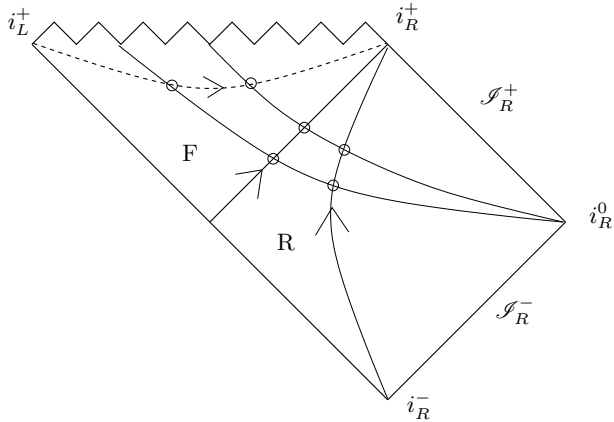}
\end{center}
\caption{
\label{fig3}
The same spacetime diagram, schematically showing a Killing slicing
that ends at the future singularity, such as Kerr-Schild slices. The
lines with arrows are trajectories of the Killing vector (lines of
constant $R$) and the beads on them represent surfaces of constant
coordinate $r$
if the Killing shift is used. In particular, the dashed line could
serve as a Killing excision boundary.
}
\end{figure}

\begin{figure}
\begin{center}
\includegraphics[width=8cm]{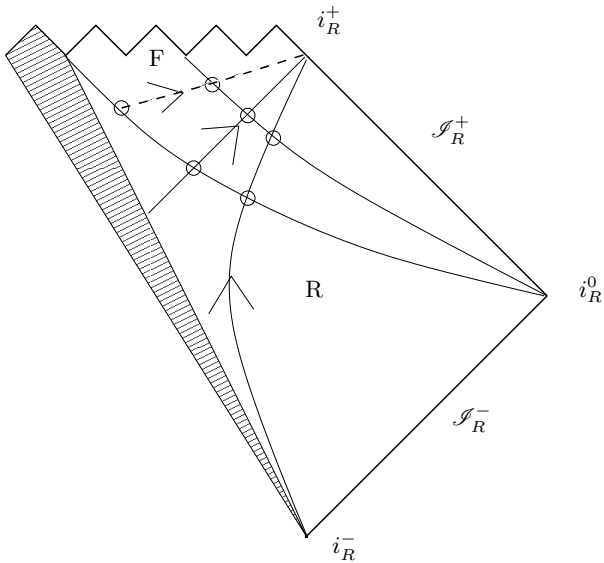}
\end{center}
\caption{
\label{fig4}
Schematic spacetime diagram of the collapse of a spherical
star. Outside the collapsing star (shaded) the spacetime is
Schwarzschild, comprising
parts of regions R and F. A Killing slicing with excision as
in Fig.~\ref{fig3} is shown. A Killing endstate cannot be reached
without excision.}
\end{figure}


\subsection{Singularity-avoiding slicings}


Both in collapse and in eternal black holes one can use slicings which
avoid the singularity. Any timelike
worldline inside a black hole has finite length, while any timelike
worldline with limited total acceleration outside the black hole has
infinite length. The lapse measures the rate of proper time per
coordinate time for an observer normal to the time slices, and so
one might think that the lapse
must go to zero everywhere inside the black hole in order to avoid the
singularity, and that because the slices keep advancing outside the
black hole, their intrinsic geometry must deform without limit as time goes on,
until large gradients can no longer be resolved. Such ``slice
stretching'' was indeed encountered in early black hole simulations,
and motivated the development of black hole excision \cite{SeidelSuen}.

Only later it was realised clearly that singularity-avoiding slicings
need not lead to slice stretching \cite{gr-qc/0206072}.  If the lapse
is chosen such that the slice is Lie-dragged along the Killing vector
field everywhere, its intrinsic geometry becomes
time-independent. This is true also inside the black hole where the
Killing vector field that generates time translations at infinity
becomes spacelike (and so the spacetime is not technically stationary),
as long as this Killing vector field is nowhere parallel to the slicing.
Once the geometry of the slice
has become time-independent, a suitable shift condition then makes the
spatial metric coefficients explicitly time-independent.  With this
lapse and shift $\partial/\partial t$ becomes the Killing vector
(spacelike inside a black hole).  Coordinate conditions which generate
Killing coordinates asymptotically starting from generic initial
coordinates were called ``symmetry-seeking'' in
\cite{GarfinkleGundlach}.

Even more recently it was realised that the lapse need not collapse
either \cite{Hannametal}. Note that
\begin{equation}
\label{dt}
\left({\partial\over\partial t}\right)^a \equiv \alpha n^a +
\beta^i\left({\partial\over\partial x^i}\right)^a
\end{equation}
is a sum of two terms. Define some scalar $\sigma$ to measure distance
from the singularity. (In Schwarzschild spacetime, an obvious choice
is the area radius $R$.) For any given $\alpha$ and $\sigma$,
$\beta^i$ then can be chosen to set $\dot\sigma=0$, except where
$\sigma_{,i}=0$.  (We use a dot to denote $\partial/\partial t$). In
other words, the lapse in a Killing coordinate system vanishes only
where the time slices are tangential to the Killing vector field.
Every regular time slice in a collapse spacetime, and every wormhole
slice through an eternal black hole has such an obstruction point,
namely a local minimum of $\sigma$ (Fig.~\ref{fig1}). However, a slice
that becomes asymptotically cylindrical (with $R\to R_0$) 
and ends at $i^+_L$ avoids
this obstruction (Fig.\ref{fig2}).


\subsection{Excision}


An alternative to singularity-avoiding slicings is singularity
excision. This means truncating the time slices along a future
spacelike surface which is also (at least asymptotically) Killing. In
Schwarzschild spacetime, this would be a surface of constant $R<2M$. One still
wants the slice to be Lie-dragged along the Killing field, but one
gains more freedom because Killing slices are now acceptable which
would intersect the singularity, such as Kerr-Schild slices of
Schwarzschild. A Killing slicing with Killing excision boundary 
is illustrated in Fig.~\ref{fig3}.

As long as the excision surface is spacelike, all characteristics
corresponding to gravitational waves, which propagate on light cones,
will be leaving the domain of computation.  Depending on the
formulation of the Einstein equations and the gauge choice, other
characteristics corresponding to constraint modes and gauge modes may
be spacelike, and either this will restrict the excision surface
further or explicit boundary conditions need to be imposed on the gauge
if the evolution equations are to be well-posed. If the system is not
hyperbolic, for example because the gauge conditions are parabolic or
elliptic, boundary conditions will be required on any excision boundary. 

In \cite{gr-qc/0411137} and \cite{gr-qc/0411149}, evolutions were
carried out from puncture data using BMt 1+log slicing with and
without $K_0$, and directly comparing evolutions using either excision
or fixed punctures. No explicit boundary condition was imposed at the
excision boundary. Excised and non-excised evolutions
are claimed to converge to each other to second order everywhere
outside the excised region. This is surprising given that the excision
problem was ill-posed. 


\section{Killing coordinates} 
\label{section:Killingcoordinates}



\subsection{General case}


By definition, coordinates in which the 4-metric is time-independent
are those in which $(\partial/\partial t)^a=C\xi^a$, where $\xi^a$ is a
Killing vector that is timelike at infinity and $C\ne0$ is a
constant. Contracting with $n_a$, we find that the {\em Killing lapse}
is given by
\begin{equation}
\label{Killinglapse}
\alpha=C\phi,
\end{equation}
where $\phi\equiv -n_a\xi^a$,
and contracting with the projector ${\perp_a}^b\equiv {g_a}^b+n_a n^b$ we
find that the {\em Killing shift} is 
\begin{equation}
\label{Killingshift}
\beta^i = {\alpha\over\phi}(\perp\xi)^i.
\end{equation}


\subsection{Schwarzschild spacetime in spherical symmetry} 


We now restrict to spherically symmetric Killing coordinate systems on
the Kruskal extension of Schwarzschild spacetime. In the following,
$X^\mu$ are preferred coordinates on a given spacetime such as
Schwarzschild, while $(t,x^i)$ are the coordinates used for the numerical
evolution, in our case with the spherical line element
\begin{eqnarray}
\label{rt}
ds^2=-\alpha^2\,dt^2+\gamma(dr+\beta\,dt)^2
+R^2\,d\Omega^2,
\end{eqnarray}
We use the shorthands $d\Omega^2\equiv
d\theta^2+\sin^2\theta\,d\varphi^2$, $R^2\equiv
\gamma_{\theta\theta}$, $\gamma\equiv\gamma_{rr}$ and $\beta\equiv
\beta^r$. We use $\dot f$ and $f'$ for the partial
derivatives with respect to $t$ and $r$. 

We use preferred coordinates $(T,R)$ on Schwarzschild with the
property that $R$ is the area radius and the Killing vector
is $\partial/\partial T$, normalised to unity at infinity, for example
Schwarzschild or Kerr-Schild coordinates. In all such coordinates
$g_{TT}=1-2M/R$ and $g_{TT}g_{RR}-g_{TR}^2=-1$.The generic Killing
coordinate system $(t,r)$ with $C=1$ is then given by the ansatz
\begin{equation}\label{KillingSlices}
T=t+F(r), \quad R=R(r).
\end{equation}
If we are interested only in the slicing, we can fix the spatial
coordinate $r$ for convenience. A better choice than using $R$ itself
as a coordinate is to make $r$ proper distance along the slice, so
that $\gamma=1$. (We shall also use the symbol $l$ for proper
radial distance.) The Killing lapse and shift
are
\begin{eqnarray}
\label{pdg1}
\alpha&=&R', \\
\label{pdg2}
\beta&=&\sqrt{\alpha^2-1+{2M\over R}}. 
\end{eqnarray}
The trace of the extrinsic curvature of the Killing slices is 
\begin{equation}
\label{pdg3}
K = 2{\beta\over R}+{\beta'\over R'},
\end{equation}
where $\beta$ is given by (\ref{pdg2}). 


\section{Evolved slicing conditions} 
\label{section:gaugeconditions}


We focus on the family of slicing conditions suggested by Bona and
Mass\'o \cite{BonaMasso} (from now BM)
\begin{equation}
\label{BMn}
\alpha\,  n^a \nabla_a \alpha \equiv \dot \alpha -\beta^i
\alpha_{,i} = -\mu_L \alpha^2 K,
\end{equation}
where $K$ is the trace of the extrinsic curvature of the slice and
$n^a$ its unit normal vector. Typically, $\mu_L>0$ is understood to be a
given function $\mu_L(\alpha)$ of the lapse. As $n^a$ is a true vector
and $\alpha$ and $K$ are scalars under a change of coordinates $x^i$
on the slice, this slicing condition is independent of the coordinates
on the slice and therefore independent of the shift. 

Confusingly, the very different slicing condition 
\begin{equation}
\label{BMt}
\dot \alpha = -\mu_L \alpha^2 (K-K_0),
\end{equation}
where $K_0(x^i)$ is the initial value of $K$
\cite{gr-qc/0206072}, is also referred to as Bona-Mass\'o
slicing. For clarity, we shall refer to (\ref{BMn}) as ``BMn'' (the
derivative is along the slice normals) and to (\ref{BMt}) as ``BMt''
(the derivative is along the time lines). 

A third slicing condition \cite{gr-qc/0008067},
\begin{equation}
\label{BMgBM}
\dot \alpha = -\mu_L \alpha\left(\alpha K - D_i\beta^i\right)
\equiv \mu_L {\alpha\over 2}(\ln|\det\gamma|)\dot{}, 
\end{equation}
where $D_i$ is the covariant derivative compatible with the 3-metric
$\gamma_{ij}$, is also related to BM. We shall call it ``BMg'', as it
can be integrated for any $\mu_L=\mu_L(\alpha)$ to relate $\alpha$ to
the 3-metric determinant. For $\mu_L=2/\alpha$, BMg integrates to
$\alpha=f(x)+\ln|\det\gamma_{ij}|$, explaining the name ``1+log
slicing''. BMn and BMt can be integrated only if the shift is zero.

The geometric specification of BMt and BMg (but not BMn)
slicing depends on the shift. Here we shall use the
``fn-driver''
\begin{equation}
\label{fn-driver}
\dot\beta^i-\beta^j\beta^i_{,j}=\mu_S\alpha^2(f^i-f^i_0),
\end{equation}
or the ``ft-driver''
\begin{equation}
\label{ft-driver}
\dot\beta^i=\mu_S\alpha^2(f^i-f^i_0),
\end{equation}
where $f^i$ is the 3-vector defined by
\begin{equation}
\label{fdef}
f_i\equiv\gamma^{jk}\gamma_{ij,k}-{\rho\over 2}
\gamma^{jk}\gamma_{jk,i}
\end{equation}
in preferred Cartesian coordinates (see Appendix~\ref{Appendix:Reduction}).
With $\rho=2/3$, these are essentially versions of the (implicit)
``Gamma-driver'' shift conditions that now dominate numerical
relativity. 

A simple analysis of BMg as a pure gauge system (similar to
Appendix~\ref{appendix:puregauge}) on Minkowski spacetime shows that
it is well-posed with a fixed shift (see also \cite{gr-qc/0303069}),
but is ill-posed with the fn or ft drivers. We do not consider it
further.

In Appendix~\ref{appendix:puregauge} we also show that BMt slicing in
combination with any shift condition always has both positive and
negative gauge coordinate speeds. This means that on any excision
surface of constant radial coordinate $r$ there will always be a gauge
mode travelling towards increasing $r$, and so excision is not
possible with this slicing condition unless a boundary condition
is imposed on the gauge at the excision boundary. A similar result
holds for the ft-driver shift condition. We will mainly use either an
algebraic Killing shift (area freezing shift) or the fn driver shift. 


\section{Compatibility of Killing coordinates with BMn slicing}
\label{section:Killingcompatible}



\subsection{General} 


In this section we ask if Killing coordinates exist that are
compatible with BMn slicing. Although the BMn slicing
condition is geometrically independent of the shift, $\alpha(x^i,t)$
only becomes time-independent if the slicing is a Killing slicing and
the shift is a Killing shift. Substituting $\dot\alpha=0$,
(\ref{Killinglapse}) and (\ref{Killingshift}) into (\ref{BMn}), we find
the scalar equation 
\begin{equation}
\label{3DHannam1}
\perp\xi^i\phi_{,i}=\mu_L(C\phi)\phi^2 K
\end{equation}
on the slice.  We use
the definitions $(\perp\xi)^a=\xi^a-\phi n^a$, $\nabla_{(a}\xi_{b)}=0$
and $K_{ab}=-\perp \nabla_a n_b$ to rewrite this equation 
as a partial differential
equation for embedding a slice with unit normal vector $n^a$:
\begin{equation}
\label{3DHannam3}
Q^{ab} \nabla_a n_b
-{1\over 2}n^a\nabla_a\psi=0,
\end{equation}
where $\psi\equiv -\xi_a \xi^a$ is related to the gravitational
potential in a stationary spacetime and
\begin{equation}
Q^{ab}\equiv -\perp\!\xi^a\!\perp\!\xi^b
+\mu_L(C\phi)\phi^2\!\perp^{ab}
\end{equation}
is a symmetric tensor intrinsic to the slice.

Given that the unit normal vector of a slice $t={\rm const.}$ is given by
\begin{equation}
n_a=-\alpha\nabla_a t, \qquad 
\alpha = (-\nabla_b t\nabla^b t)^{-1/2}, 
\end{equation}
the principal part of (\ref{3DHannam3}) is $Q^{ab}\nabla_a\nabla_b t$.
As $\perp^{ab}$ is positive definite and $\mu_L>0$, two eigenvalues of
$Q^{ab}$ are always positive. The third eigenvalue is associated with
the eigenvector $\perp\!\xi^a$ and is given by
\begin{equation}
\label{Ddef}
D=(\mu_L-1)\phi^2+\psi.
\end{equation}
Therefore (\ref{3DHannam3}) is elliptic for $D>0$ and (2+1) hyperbolic
for $D<0$.

Alcubierre \cite{gr-qc/0210050} has shown that the BMn slicing
condition can also be written as the 3+1 wave equation
\begin{equation}
\label{Alcubierreslicing}
P^{ab}\nabla_a \nabla_b t =0, \qquad
P^{ab}\equiv -n^a n^b +\mu_L(\alpha)\perp^{ab}
\end{equation}
where $\perp^{ab}$, $\alpha$ and $n^a$ are as given above.  We have
perturbed this equation around a Killing slicing $t$, but have not been
able to identify any lower-order (friction-like) terms that would
always push $\delta t$ locally towards $\xi^a\nabla_a \delta t=0$ or
$\delta t=0$. We conclude that if BMn slicing is really symmetry
seeking in some circumstances, as our numerical evidence below
suggests, this is not because of local friction terms, but rather
through the mechanism by which a solution of the wave equation on a
finite domain with a dissipative boundary condition settles to a
time-independent solution of the Laplace equation. 

The characteristics of the wave equation
(\ref{Alcubierreslicing}) are null surfaces of the ``gauge metric''
$(P^{-1})_{ab}=-n_an_b+\mu_L^{-1}\!\!\perp_{ab}$, which is the matrix
inverse of $P^{ab}$. A slice evolving under (\ref{Alcubierreslicing})
can be excised on a boundary ruled by trajectories
of the Killing vector only if the Killing vector is ``spacelike''
with respect to the gauge metric, that is
$(P^{-1})_{ab}\xi^a\xi^b>0$. We find that this is once again equivalent to
$D<0$. 


\subsection{Schwarzschild spacetime in spherical symmetry}
\label{section:sphericalKilling}


This subsection reviews and generalises \cite{Hannametal}.
The BMn Killing slicing condition in spherical symmetry is
\begin{equation}
\label{2ndorderODE}
\beta\alpha'=\mu_L(\alpha)\alpha^2 K
\end{equation}
Using (\ref{pdg1}) and (\ref{pdg3}) to eliminate $\alpha$ and $K$
gives 
\begin{equation}
\label{tmp1}
-{R''\over R'\,\mu_L(R')}+{\beta'\over\beta}+2{R'\over R}=0, 
\end{equation}
which has an obvious first integral that can be expressed, using (\ref{pdg1})
and (\ref{pdg2}), as
\begin{equation}
\label{firstintegral}
-2\int^{R'}{d\alpha\over\alpha\,\mu_L(\alpha)}
+\ln{\left[\left(R'^2-1+{2M\over R}\right) R^4\right]}=c.
\end{equation}
Alternatively, using (\ref{pdg2}) to eliminate $\beta$ from (\ref{tmp1}) gives
\begin{equation}
\label{HannamODE}
R''=-{\mu_L\over R} {N \over D},
\end{equation}
where 
\begin{eqnarray}
N&\equiv&R'^2 \left(2R'^2 - 2 + {3M\over R}\right), \\
D&\equiv&[\mu_L(R')-1]R'^2 +1 - {2M\over R}.
\end{eqnarray}
[Here $D$ has the same meaning as in (\ref{Ddef}).]  For given
$\mu_L(\alpha)$ this is a second order ODE for $R(r)$. For the
solution to be regular for all $R>0$, $N$ and $D$ have to vanish at
the same $r$, which becomes a regular singular point. This fixes $R$
and $R'$ at this $r$, and hence the constant $c$ in
(\ref{firstintegral}). (\ref{firstintegral}) can then be solved as a
first-order ODE for $R(r)$. This means that for any $\mu_L(\alpha)$,
there are at most a finite number of twice differentiable spherically
symmetric Killing slicings of Schwarzschild, one for each possible
regular singular point. The 3-dimensional PDE (\ref{3DHannam3}) of
which (\ref{HannamODE}) is the reduction to spherical symmetry is
elliptic for $R>R_c$ and hyperbolic for $R<R_c$. In the absence of
spherical symmetry, requiring regularity at the 2-dimensional boundary
between elliptic and hyperbolic regions would also make the slice more
rigid, as it does in spherical symmetry. The first integral
(\ref{firstintegral}), however, has no counterpart in the absence of
spherical symmetry.

\paragraph*{1+log slicing}

The case of 1+log slicing, $\mu_L=2/\alpha$ has been presented in
\cite{Hannametal}, based on earlier work in
\cite{EstabrookWahlquist}. There are two possibilities for regular
singular points. One is $R=2M$ with $R'=0$. This gives a Killing
slicing where each slice goes through the bifurcation point of the
horizon, the lapse is positive in R and negative in L, and the slices
never reach P or F. It is not of interest for numerical
evolutions. 

The other regular singular point is $R'=R_c'\equiv -3+\sqrt{10}$,
$R=R_c\equiv M/(4R_c')\simeq 1.54057 M$. In this solution $R\to\infty$
as $r\to\infty$ and $R\to R_0$ from above as $r\to-\infty$. $R_0$ can be found
from (\ref{firstintegral}) with $R'=0$, is given in implicit form in
\cite{Hannametal}, and is approximately $R_0\simeq 1.31241 M$. Inside
the black hole the slices become asymptotically tangent to the Killing
field and terminate at $i^+_L$. The intrinsic geometry of each slice
becomes a cylinder of radius $R_0$ as
$r\to-\infty$ (Fig.~\ref{fig2}).

\paragraph*{Harmonic slicing}

Harmonic slicing is the special case of BMn slicing with
$\mu_L=1$. The regular singular points are then $R=2M$ with either
$R'=0$ or $R'=\pm 1/2$. The former can be discarded, and the sign in
the latter is trivial, so that the Killing slices are characterised by
$\alpha=R'=1/2$ at $R=2M$. These slices stretch from $i^0_R$ to the
future singularity $R=0$, and so must be used with excision. The gauge
characteristics are the light cones \cite{bssn3}, so the gauge only
requires the excision boundary to be spacelike. 

\paragraph*{General $\mu_L(\alpha)$}

Killing slices cannot have an extremum of $R$ if they are to be
stationary points of some slicing condition. From
(\ref{firstintegral}) we see that if the Killing slices are to
approach $i^+_L$, that is $\lim_{r\to\infty} R =R_0>0$, the integral
\begin{equation}
\int^0{d\alpha\over\alpha\,\mu_L(\alpha)}
\end{equation}
must be finite, for example with $\mu_L=2/\alpha$. We conjecture that,
conversely, if this integral diverges, as with $\mu_L=1$, the
Killing slices must intersect the future singularity. 

\paragraph*{Excision and uniqueness}

One might think that excising a BMn Killing slice would make it less
rigid, because the regular singular point $R=R_c$ could be
excised. This is correct if one excises at $R_c<R<2M$ and imposes an
explicit boundary condition on the slicing, for example by fixing
$\alpha$ at the excision boundary. By function counting one would
expect the value of $\alpha$ at the boundary to control the value of
the constant $c$ of the slice.  However, to excise all modes including
the lapse gauge modes, the excision boundary must be in the region
where $D<0$, and so $R=R_c$ must be on the slice. The only possible
Killing endstate of the slicing is then the unique one derived above.


\section{Vacuum black hole evolutions}
\label{section:vacuumevolutions}



\subsection{Method}


To see empirically if generic black hole evolutions are attracted to
the Killing states we have characterised above, we have carried out numerical
evolutions of the Schwarzschild spacetime in spherical symmetry, using
BMn 1+log slicing.

We can take advantage of the fact that this metric is known in
closed form to evolve only the coordinates on the known spacetime, see
Appendix~\ref{appendix:puregauge}. There is no global coordinate
system that covers wormhole slices and which is also
Killing. Therefore, in pure gauge evolutions of wormhole slices
stretching from $i^0_R$ to $i^0_L$, we restrict to slices with a
discrete ``left-right'' isometry through the coordinate sphere $r=0$,
so that we only evolve explicitly on F and R, where KS coordinates can
be used, with a boundary condition at $r=0$ representing the
isometry. 

Even this does not work for slices which go through the horizon
bifurcation 2-sphere (where KS time and similar Killing time
coordinates are $-\infty$), and so for such slices we need to evolve
the Einstein equations in the NOR formulation, see
Appendix~\ref{appendix:wormhole}. In all other cases, plots are from
pure gauge evolutions, but we have verified that our results are
replicated in evolutions of the full Einstein equations in the NOR
formulation. The evolutions described here all use the fn shift condition
(\ref{fn-driver}) except otherwise stated.


\subsection{With excision boundary}


As initial data for the geometry and the coordinates we
have considered:

1a) KS slice, KS lapse, KS shift, area radius;

1b) KS slice, KS lapse, zero shift, area radius;

2) A closed form asymptotically cylindrical slice, unit lapse, zero
   shift, area radius, see Appendix~\ref{appendix:gaugedata}.

3) The Hannam slice, lapse, shift, all in area radius, see
Appendix~\ref{appendix:gaugedata}.

\begin{figure}
\begin{center}
\includegraphics[width=0.45\textwidth]{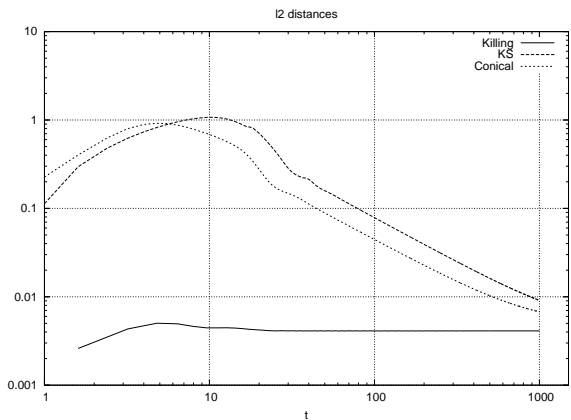}
\caption{\label{Fig:endstate} The $ L^2 $ distance of the lapse from
the Killing endstate over the range from the excision boundary
$R=1.54$ (just inside the regular singular point) out to $R=21.54$,
with area locking shift. The power law decay indicates
$||\alpha-\alpha_ {{\rm Killing}}||\sim t^{-1}$.}
\end{center}
\end{figure}

We first evolved with area locking (that is, Killing) shift (which is
determined algebraically so that the initial value of the shift listed
above is irrelevant) and excision.  We excised at $R=1.54M$, which is
just inside the maximal excision radius $R=R_c\simeq 1.54057M$ for
which all modes are outgoing.  We find that 1) and 2) approach the
Killing state, and 3) remains there. This is demonstrated in
Fig.~\ref{Fig:endstate}, and indicates that the Killing state has a
significant basin of attraction.

When combined with the fn shift driver, in 1a) the coordinates $r$ are
pushed out of the black hole and further. 1b), and 2) again settle
down to the Hannam endstate, and 3) remains there. With 1b), the
excision radius initially has to be $R\simeq 1.3M$ or the excision
surface at constant $r$ will be pushed out so far before it reaches
steady state that there a gauge mode is ingoing.

\begin{figure}
\begin{center}
\includegraphics[width=0.45\textwidth]{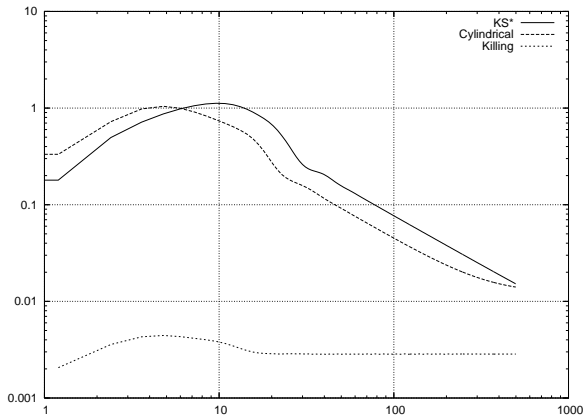}
\caption{\label{Fig:endstateshift} The $ L^2 $ distance of the
lapse from the Killing endstate from the excision boundary
with approximately the same limits as in Fig.~\ref{Fig:endstate},
using the fn shift driver (\ref{fn-driver}). 
The power law decay again indicates
$||\alpha-\alpha_{{\rm Killing}}||\sim t^{-1}$.}
\end{center}
\end{figure}


\subsection{With isometry boundary condition}


We begin with NOR evolutions starting from the time-symmetric wormhole
slice that goes through the bifurcation 2-sphere $R=2M$. 

We first use spatial coordinates in which $i^0_L$ is represented by
the point $r=0$ (the ``puncture''), see
Appendix~\ref{appendix:wormhole}. Evolutions with this method reproduce
the behaviour described in \cite{Hannametal}: numerical error changes
the topology and the evolution settles down to the asymptotically
cylindrical Killing state.

If we evolve the same initial data in spatial coordinates that
resolve the wormhole (see Appendix~\ref{appendix:wormhole}), we see
the slices begin to form a cylinder at radius $R_0$, but at reasonable
resolution constraint violation in the Einstein code makes the result
unreliable soon after.

\begin{figure}
\begin{center}
\includegraphics[width=0.45\textwidth]{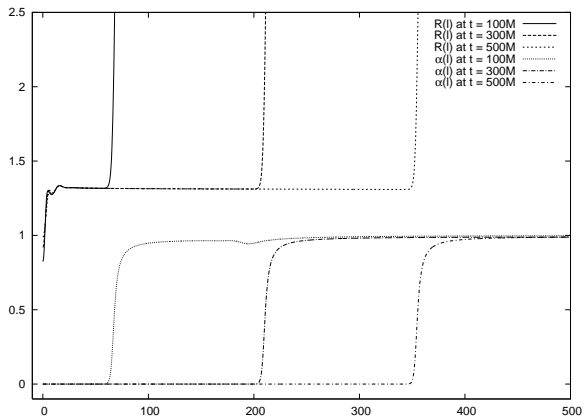}
\caption{\label{Fig:cylinder} Snapshots of $R$
and $\alpha$ against proper distance $r$ from an evolution of an 
isometric slice. The throat of the slice (initially at $R=1.5M$)
is gradually stretched so that it becomes an infinitely long 
cylinder. The radius of the cylinder agrees with that computed 
in \cite{Hannametal}. Note that low numerical resolution effectively
smears out a gauge shock travelling left, so that this is not a
correct continuum solution. }
\end{center}
\end{figure}

Pure gauge evolutions with wormhole initial data that lie to the
future of the bifurcation 2-sphere (so that $R<2M$ at the throat) and
a discrete isometry boundary as described in
Appendix~\ref{appendix:puregauge} are more stable. At the
isometry boundary (where $R$ is minimal) the lapse quickly collapses.
In low resolution evolutions, the lapse collapses starting at the
minimal $R$ (at the isometry boundary), and a cylinder of radius $R_0$
forms with proper length increasing linearly in
time (Fig.~\ref{Fig:cylinder}).


\subsection{Gauge shocks}


However, higher resolution (for example $\Delta r=M/50$) evolutions
show that low resolution only hides the formation of a gauge shock
where $K$ forms a large negative peak and $\alpha''$ forms a positive
peak, at $R\simeq 1.5M$. This does not seem to happen exactly at $R_c$
(we varied $\mu_L(\alpha)$ to check this), and so we do not think that
it is a kink instability related to the regular singular point of the
Hannam slice. Neither is there any indication that the slice has
become null. An ODE mechanism by which $K<0$ makes $\alpha$ grow is
also ruled out as not all initial data where $K<0$ shock.

Rather, we think we see a gauge shock of the type described by
Alcubierre \cite{gr-qc/9609015,gr-qc/0210050}. Note that the lapse
speeds expressed in terms of proper distance $l$ per coordinate time
$t$, relative to the time lines, are $-\beta\pm
\alpha\sqrt{\mu_L}=-\beta\pm \sqrt{2\alpha}$, so that a gauge wave
propagating ``left'', from high to low $\alpha$ is expected to
steepen. By contrast, the wave propagating ``right'' and forming the
cylinder appears to be stable and translating with constant speed
$dl/dt$ without changing its shape much. Alcubierre notes that for the
particular choice $\mu_L=1+k/\alpha^2$ with $k>0$ the pure gauge
system is linearly degenerate, and we have tried this $\mu_L$, but
shocks still form, also in agreement with Alcubierre's numerical
observations. Alcubierre argues that gauge shocks are generic for
evolved gauge conditions. 

\begin{figure}
\begin{center}
\includegraphics[width=0.45\textwidth]{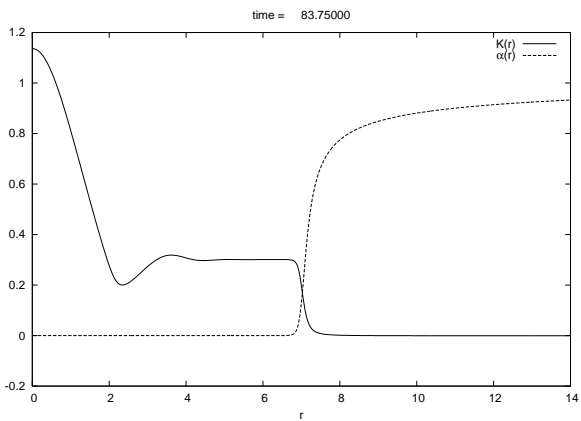}
\caption{\label{Fig:BMnNoKink} The $K=0$ time symmetric slice through
the bifurcation surface of Schwarzschild, evolved with BMn 1+log
slicing, with $\alpha=1$ initially. We show a snapshot of $K$ and
$\alpha$ against proper distance radius. The edge at $r\simeq 7$ in this
snapshot moves to the right, and leaves behind a cylinder of constant
$R$ and $K$ with $\alpha\simeq 0$. }
\end{center}
\end{figure}

\begin{figure}
\begin{center}
\includegraphics[width=0.45\textwidth]{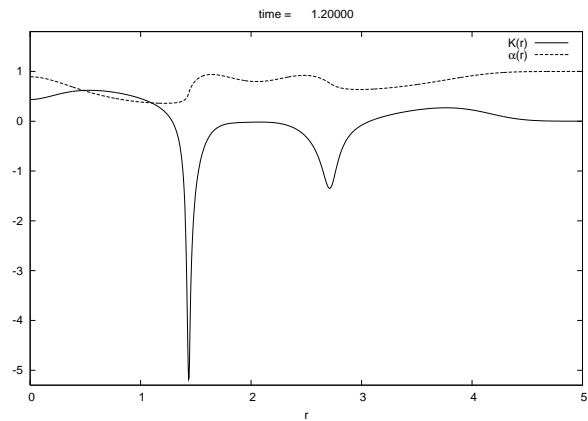}
\caption{\label{Fig:BMnKink} As in Fig.~\ref{Fig:BMnNoKink}, but with
$\alpha$ not constant on the initial slice. The wave on the left
  travels left and is steepening, about to form a gauge shock, with
  large negative $K$. The wave on the right travels right and is also
  steepening: note $\alpha''>0$ there. }
\end{center}
\end{figure}

Although the NOR evolutions of the time-symmetric slice are less
reliable, they suggest that evolutions shock when $\alpha$ has a local
minimum not at the isometry boundary (Fig.~\ref{Fig:BMnKink}). They
also suggest that with $\alpha=1$ initially the slicing never shocks
(Fig.~\ref{Fig:BMnNoKink}).  This agrees with the standard numerical
literature where the puncture data are approximately the
time-symmetric slice through Schwarzschild and the initial lapse is
one. It also agrees with the evolution by Brown \cite{Brown} of these
particular initial data. It seem plausible that initial data in a
neighbourhood also do not develop shocks, but we have not investigated
this. 

We note that the shift remains regular during the blow-up, and the
same qualitative picture occurs with proper distance radius, zero
shift, or fn driver shift.

With BMt slicing and the ft (not fn) shift driver, we see the same gauge
shock in both NOR and pure gauge evolutions, but it seems to form
earlier and even at low resolution, so that we never see formation of
a cylinder before the code crashes. 

\section{Scalar field collapse evolutions}
\label{section:scalarfieldcollapse}


We now consider the behaviour of collapse simulations with BMn 1+log
slicing. As a toy model we consider a spherical scalar field.  We
impose spherical symmetry and use proper distance as the radial
coordinate.  The metric thus takes the form (\ref{rt}) with
$\gamma=1$. Details of the numerical implementation and the initial data
are given in Appendix~\ref{section:collapsecode}.

\begin{figure}
\includegraphics[width=0.45\textwidth]{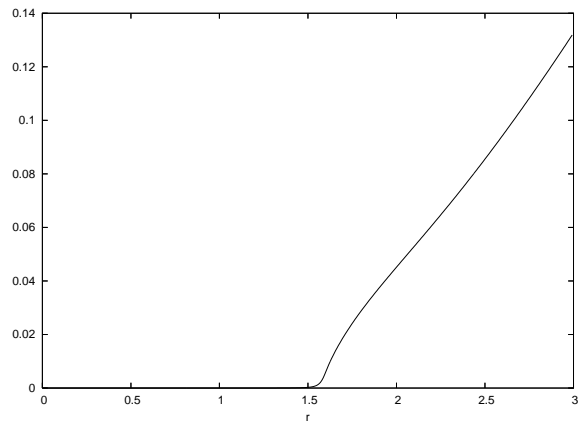}
\caption{\label{alphanoex}Plot of the lapse $\alpha$ against proper
distance $r$ at $t=14$ in scalar field collapse without excision.}
\end{figure}

\begin{figure}
\includegraphics[width=0.45\textwidth]{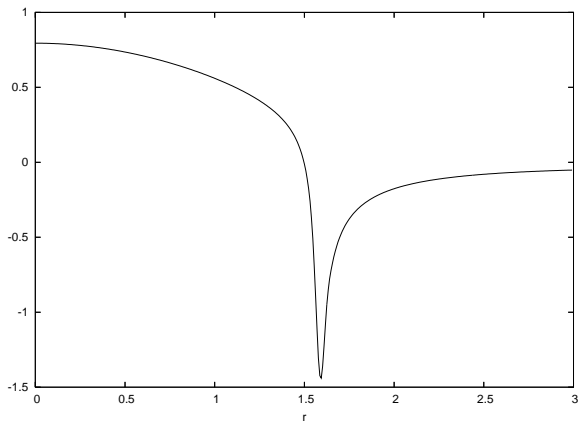}
\caption{\label{knoex}Plot of $K$ against proper distance $r$ at
$t=14$ without excision.}
\end{figure}

\begin{figure}
\includegraphics[width=0.45\textwidth]{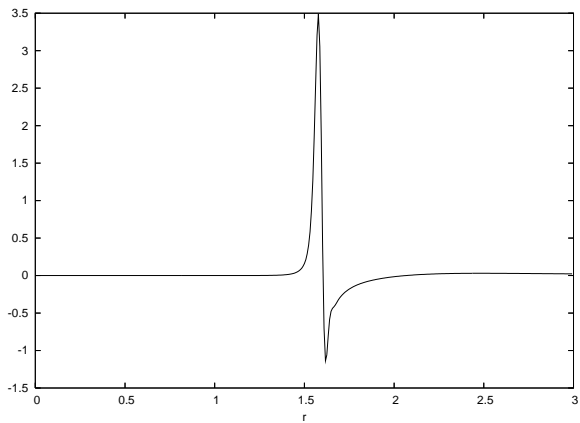}
\caption{\label{d2alphanoex}Plot of $\alpha''$ against distance $r$ at
$t=14$ without excision.}
\end{figure}

The initial data are chosen as a moment of time symmetry. The scalar
field separates into an ingoing pulse and an outgoing pulse. With the
chosen parameters, the ingoing pulse collapses to form a black hole,
with an apparent horizon first forming at $t=6.9$.  The final mass of
the black hole is $1.0$. Figs.~\ref{alphanoex} and \ref{knoex} show
respectively $\alpha$ and $K$ at $t=14$ and in the range $0\le r \le
3$.  Note the sharp features in both these quantities near $r=1.5$.
These features become ever sharper and cause the code to crash not
long after the time of these graphs.  This pathology is again a gauge
shock.  Neglecting the shift the principal part of the the evolution
equation for $K$ is ${\dot K} = - {\alpha ''}.$ Combining this with
the BMn equation yields a nonlinear wave equation for the lapse whose
principal part is ${\ddot \alpha} = 2 \alpha {\alpha ''}.$ The modes
of this equation travel with speeds $\pm {\sqrt {2 \alpha}}$.  Thus if
one has an inner region where the lapse has collapsed, then
left-moving gauge waves pile up on the boundary of this region, giving
rise to a shock wave in $\alpha$ which in turn (through the equation
${\dot K} = - {\alpha ''}$) will induce a shock wave in $K$.
Fig.~\ref{d2alphanoex} shows $\alpha ''$ at $t=14$ and in the range
$0\le r \le 3$.  Note that this quantity also has a sharp feature near
$r=1.5$.

\begin{figure}
\includegraphics[width=0.45\textwidth]{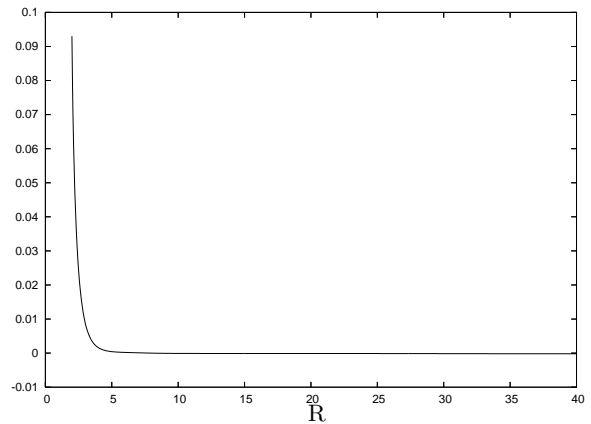}
\caption{\label{kexcise}Plot of $K$ against area radius $R$ at
$t=60.5$ with excision.}
\end{figure}

\begin{figure}
\includegraphics[width=0.45\textwidth]{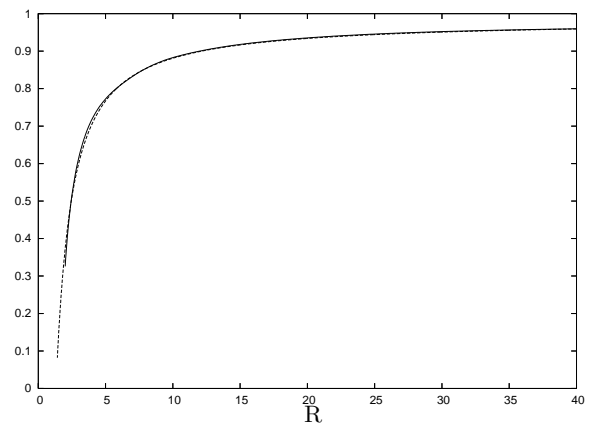}
\caption{\label{alphacompare}Plot of $\alpha$ against area radius $R$
at $t=60.5$ with excision (solid line) and the exact Killing lapse
(dashed line).}
\end{figure}

In this simulation, the pathological behaviour is inside the horizon.
This suggests that we might be able to avoid the pathology by using
excision.  Figs.~\ref{alphacompare} and \ref{kexcise} respectively
present the values of $\alpha$ and $K$ for a simulation done using
excision.  Here the simulation is run until $t=60.5$ and all
quantities are plotted as functions of the area radius $R$ rather than
proper distance $r$.  It is excision that allows the simulation to be
run this long because the excised grid contains no regions of negative
$K$ which caused the non-excision simulation to crash. Furthermore,
this late in the simulation, these quantities have asymptoted to the
static values described in \cite{Hannametal}.  This is illustrated in
Fig.~\ref{alphacompare} which contains two plots of $\alpha$ as a
function of $R$. The solid line is the $\alpha$ given by the endstate
of the excision collapse simulation, while the dotted line represents
the Killing lapse given by an integration of the ODEs of \cite{Hannametal}.
 
\section{Conclusions}
\label{section:conclusions}


\paragraph*{Killing endstates}

We have explained why it is possible in evolutions of black holes that
all metric coefficients become time-independent without either slice
stretching or collapse of the lapse. We have reviewed the Bona-Mass\'o
slicing conditions, and have derived a mixed elliptic/hyperbolic PDE
on the slice that characterises Killing endstates of the BMn family of
slicing conditions. Numerically, we have shown that spherical BMn
slicings of the Schwarzschild spacetime are attracted to the Killing
endstate from nearby initial data. We do not fully understand the
mechanism for this. Initial data further away also appear to be
attracted to the Killing state, but on closer inspection this is true
only at low numerical resolution. 

\paragraph*{Gauge shocks}

Increasing the resolution reveals that in the continuum the 1+log BMn
slicing {\em generically} develops gauge shocks of the type described
by Alcubierre \cite{gr-qc/9609015}, where the speed of gauge waves
associated with the slicing increases with the lapse, so that gauge
waves moving from large to small lapse steepen. The only initial data
set we have examined that does {\em not} form a gauge shock with BMn
1+log slicing is the time-symmetric wormhole slice through
Schwarzschild spacetime with unit initial lapse, although we suspect
that there is at least a neighbourhood of such data. More numerical
work is required to explore this.

It may be that gauge shocks would also occur in binary black hole
simulations with 1+log BMn slicing in the continuum, but that they are
suppressed by low resolution inside the black holes. By contrast, in
collapse simulations the central region is typically adequately
resolved, and in fact recent work where the
collapsing region is never excised seems to require large dissipation for
stability \cite{BaiottiRezzolla}.

\paragraph*{Excision}

We find that in both collapse and vacuum simulations the gauge shock can
typically be avoided by excising just inside the apparent
horizon. There seems to be no clear awareness in the literature that
such a boundary still has an incoming gauge mode, and that the
resulting continuum problem is ill-posed. 

Confirming this, our vacuum
(both Einstein and pure gauge) evolutions in spherical symmetry do not
converge and often blow up when an incoming mode at the excision
boundary is neglected. By contrast, our collapse code, which uses
different numerical methods, does not seem to mind. There is also an
explicit claim that 3D binary black hole evolutions converge equally
well with and without excision \cite{gr-qc/0411137,gr-qc/0411149}.

We have also re-derived the previously known \cite{bssn3} fact that if
gauge drivers are not of the form $\dot\alpha +
\beta^i\alpha_{,i}=\dots$ and $\dot
\beta^i+\beta^j\beta^i_{,j}=\dots$, full excision is not possible at
any radius.

\paragraph*{Nature of the Killing endstate of BMn slicing}

If there is to be no incoming gauge mode at an excision boundary, the
equation obeyed by the Killing slice has a transition from elliptic to
hyperbolic. Requiring regularity there makes the slice more rigid, and
in spherical symmetry makes it unique. The same is true if the slice
has no excision boundary.

We have clarified that this unique Killing BMn slicing of a
Schwarzschild black connects spacelike infinity outside the black hole
to future timelike infinity inside the black hole, where it becomes
asymptotically cylindrical. As pointed out independently by Brown
\cite{Brown}, initial data which connect two asymptotically flat
regions through a wormhole cannot evolve to this endstate in the
continuum, although numerical under-resolution gives the false
impression that the topology jumps. In the continuum evolution, the
wormhole stretches into a cylinder whose length grows without bound.

\paragraph*{Comments on 3D evolutions}

In our investigation, we have identified three problems with current
gauge choices in 3D numerical evolutions of collapse and black holes
with a currently favoured slicing condition, BMn 1+log slicing: 1)
wormhole data do not admit a BMn Killing endstate; 2) excision close
inside the apparent horizon requires explicit boundary conditions for
the gauge; and 3) coordinate shocks form generically.  None of these
problems have been noted in the binary black hole literature, but we
believe that this is only because of limited resolution, and that they
will become apparent as a failure of convergence or instabilities at
sufficiently high resolution. There are, however, simple ways around
these problems:

\begin{itemize}

\item Wormhole initial data for eternal black holes ending at $i^0_L$
should be replaced by initial data that are asymptotically cylindrical
and end at $i^+_L$. 

\item Continuum boundary conditions should be imposed explicitly at
excision boundaries for any incoming gauge modes. 

\item The initial lapse should be chosen such that gauge shocks do not
form. This will require more empirical studies in 3D. In collapse
without excision, changing to a smoothly collapsed lapse profile once
an apparent horizon has formed may be helpful.

\end{itemize}

\paragraph*{Final remarks}

Finally, two technical developments given in the appendix may also be
of interest to the 3D community. By characterising pure gauge as the
evolution of a coordinate system on a background spacetime given as
$X^\mu=X^\mu(x^i,t)$, we have been able to check strong hyperbolicity
of the gauge and calculate the gauge speeds without reference to a
formulation of the Einstein equations. By re-defining the vector
auxiliary variable of the NOR and BSSN formulations (following
\cite{Meudonharmonic}), we have made them easier to use with
non-Cartesian coordinates or multiple coordinate patches.

\acknowledgements

We would like to thank Niall \'O Murchadha for helpful discussions and
Jos\'e M. Mart\'\i n-Garc\'\i a for helpful discussions and comments on the
manuscript.  DG was supported in part by NSF grant PHY-0456655 through
Oakland University.


\appendix


\section{Pure gauge evolutions}
\label{appendix:puregauge}



\subsection{General equations}


Prescriptions for the lapse and shift can be tested without the
specific stability problems associated with any formulation of the
Einstein equations by evolving coordinates $(t,x^i)$ on a spacetime
given a priori in coordinates $X^\mu$ with metric $g_{\mu\nu}$. The
ADM evolution equations are replaced by the definition (\ref{dt}) of
$\partial/\partial t$ in terms of the lapse and the shift,
\begin{equation}
\dot X^\mu = \alpha n^\mu + \beta^i X^\mu_{,i}, 
\end{equation}
where $n^\mu$ are the $X^\mu$ components of the unit normal to the
$t$-slices. The 3-metric is given by 
\begin{equation}
\gamma_{ij}=g_{\mu\nu}
X^\mu_{,i}X^\nu_{,j},
\end{equation}
and $\dot\gamma_{ij}={\cal L}_\beta \gamma_{ij}-2\alpha K_{ij}$ gives
\begin{eqnarray}
K_{ij} &=&  {X^\mu}_{,ij} n^\nu g_{\mu\nu}  + {1\over 2} \Bigl(
{X^\mu}_{,i} {X^\lambda}_{,j} n^\nu \br + {X^\lambda}_{,i} {X^\nu}_{,j} n^\mu  
- {X^\mu}_{,i} {X^\nu}_{,j} n^\lambda 
\Bigr)g_{\mu\nu,\lambda}, 
\end{eqnarray}
where the unit normal $n^\mu$ is given, up to normalisation, by
\begin{equation}
n^\mu \propto g^{\mu\nu}\epsilon_{\nu\alpha\beta\gamma}
{X^\alpha}_{,1} {X^\beta}_{,2} {X^\gamma}_{,3}.
\end{equation}

In spherical symmetry $n_a$ is defined by $(\partial/\partial r)^a
n_a=0$, and in preferred coordinates $(T,R)$
\begin{equation}
n_T = -{R'\over\sqrt{\gamma}},\quad
n_R = {T'\over\sqrt{\gamma}}.
\end{equation}


\subsection{Initial data}
\label{appendix:gaugedata}


We use KS coordinates for the background Schwarzschild spacetime. Simple
closed form slices which become asymptotically cylindrical at $R=R_0$
may be constructed by making the ansatz $T=t+F(R)$ with
\begin{eqnarray}\label{SliceAnsatz}
F'(R)&\simeq&-L(R-R_0)^{-1}.
\end{eqnarray}

Constructing 1+log Killing data is straightforward in area gauge where
$r=R$. We note that proper distance $l$ is $dl=\sqrt{\gamma}dR$
in area gauge $r=R$, while $R'=dR/dl$ in proper distance gauge
$r=l$. Therefore, we obtain an algebraic expression for
$\gamma$ in area gauge by replacing $R'$ by
$\sqrt{\gamma}$ in (\ref{firstintegral}). A single numerical
integral must be performed to compute $F(r)$.

A surface with the required isometry in the F region, with a radial
coordinate $r$ in which the isometry is $r\to-r$, can be given in
Schwarzschild coordinates $(\eta,R)$ as $R(-r)=R(r)$ and
$\eta(-r)-\eta_*=-(\eta(r)-\eta_*)$. Here $\eta=\eta_*$ is the
reflection surface. Note it is timelike in F. The last condition must
be translated into a relation between $T(-r)$ and $T(r)$ using
$\eta=T-\phi(R)$ where $\phi(R)\equiv 2M\ln(1-2M/R)$. Furthermore, we
want the Kerr-Schild time $T$ to be smooth through $R=2M$, while
$\phi(R)$ is not. An ansatz
with these properties is
\begin{eqnarray}
T(r)&=& C(r)r+[1+D(r)]\phi(R(r)),
\end{eqnarray}
where $C(r)$ and $D(r)$ are smooth odd functions and $D(r)$ 
is $0$ at $r=0$ and $-1$ on and outside the horizon. To preserve the 
isometry and the adapted radial coordinate in the evolution 
the lapse and shift must obey $\alpha(-r)=\alpha(r)$, $\beta(-r)=-\beta(r)$.


\subsection{Characteristic analysis}


The spherical gauge evolution system is not quasilinear, because the
equations for $\dot R$ and $\dot T$ are nonlinear in $R'$ and
$T'$. Hence in order to analyse the hyperbolicity of the system one
must explicitly linearise, 
and apply weights to equations and variables \cite{R&R}. 
The principal part of the evolution of the coordinates is then
\begin{eqnarray}
\dot{\delta T}&\simeq&\Bigg(\beta+\frac{\alpha}{\sqrt{\gamma}}\Big(
g_{TR}+n^Rn^T\Big)\Bigg)\delta T' \br
+\frac{\alpha}{\sqrt{\gamma}}\Bigg(g_{RR}
-(n^T)^2\Bigg)\delta R' 
+n^T\delta\alpha+ T'\delta\beta,\\
\dot{\delta R}&\simeq&\frac{\alpha}{\sqrt{\gamma}}\Bigg(g_{TT}
+(n^R)^2\Bigg)\delta T' \br
+\Bigg(\beta+\frac{\alpha}{\sqrt{\gamma}}\Big(g_
{TR}+n^Rn^T\big)\Bigg)\delta R'\br+n^R\delta\alpha+ T'\delta\beta.
\end{eqnarray}

For BMn slicing with ft shift driver, this is completed by 
\begin{eqnarray}
\dot{\delta\alpha}\simeq\frac{\alpha^2\mu_{L}}
{\gamma}(n_R\delta T''+n_T\delta R'')+\beta\delta\alpha',\\
\dot{\delta\beta}\simeq-\frac{\alpha^2(2-\rho)\mu_S}
{\gamma^{3/2}}(n^R\delta T''+n^T\delta R'').
\end{eqnarray}
The system is diagonalisable with real characteristic 
speeds $dr/dt$
\begin{eqnarray}
\label{BMnspeed}
&-\beta\pm\frac{\sqrt{\mu_L}}{\sqrt{\gamma}}
\alpha,& \\
\label{ftdriverspeed}
&\frac{1}{2}\left(\beta
\pm\sqrt{\beta^2+\frac{4\mu_S\alpha^2 (2-\rho)}{\gamma}}\right),&
\end{eqnarray}
and so is strongly hyperbolic. 

The principal part of BMn with the fn shift driver (\ref{fn-driver})
is the same except that $\dot{\delta\beta}=\dots+\beta
\delta\beta'$. This system is also strongly hyperbolic with
characteristic speeds
\begin{eqnarray}
&-\beta\pm\frac{\sqrt{\mu_L}}{\sqrt{\gamma}}
\alpha,& \\
\label{fndriverspeed}
&-\beta\pm\frac{\sqrt{\mu_S}\sqrt{2-\rho}}{\sqrt{\gamma}}\alpha.&
\end{eqnarray}

For BMn with area locking shift the principal part becomes
\begin{eqnarray}
\dot{\delta T}&\simeq&\beta\delta T'
-\frac{1}{n_T}\delta\alpha,\\
\dot \alpha &\simeq&-\frac{\alpha^2\mu_L n_T}
{\gamma}\delta T''+ \beta\delta\alpha'.
\end{eqnarray}
In this case the system is also strongly hyperbolic and has 
characteristic speeds (\ref{BMnspeed}).

The BMt slicing condition (\ref{BMt}) gives rise to strongly 
hyperbolic pure gauge systems with speeds
\begin{eqnarray}
\label{BMtspeed}
\frac{1}{2}\left(\beta
\pm\sqrt{\beta^2+\frac{4\mu_L\alpha^2}{\gamma}}\right),
\end{eqnarray}
with either area locking shift, ft shift driver
[adds speeds (\ref{ftdriverspeed})]  or fn shift driver
[adds speeds (\ref{fndriverspeed})].

Constructing boundary conditions for a system which is not quasilinear
is a difficult task which will not be discussed here. Numerically, we
have simply frozen all fields at the outer boundary, but we have moved
the outer boundary so far out that in the continuum it does not affect
the results shown here.


\section{Wormhole initial data for the Einstein equations}
\label{appendix:wormhole}


To create a numerical evolution in spherical symmetry that is similar
to the BSSN moving puncture evolutions in 3D, we use the well-known
isotropic radial coordinate
\begin{equation}
R(r)=\left(1+\frac{M}{2r}\right)^2r,
\end{equation}
with range $0<r<\infty$.  We use the correct reduction to spherical
symmetry of the NOR system described in Appendix
\ref{Appendix:Reduction}, stagger the grid around $r=0$, and impose as
boundary conditions at $r=0$ that $\gamma_{rr}$, $R$ and $\alpha$ are
even in $r$, and $f^r$ and $\beta^r$ are odd. These conditions hold
for the puncture initial data, and it is easy to see that they are
compatible with the time evolution. If $r=0$ is a regular centre of
spherical symmetry, then $(-r,\theta,\varphi)$ represents the same
point as $(r,\pi-\theta,\varphi+\pi)$, and these conditions follow
from spherical symmetry and regularity. If $r=0$ represents $i^0_L$
then $r<0$ is simply not part of the spacetime, the even/odd
conditions do not follow from spherical symmetry, and their meaning is
unclear. However, for finite differencing purposes they are equivalent
to finite differencing across the puncture $x=y=z=0$ in 3D {\em as if} it
was a regular point, which is what is done in 3D puncture evolutions.

In order to resolve both sides of the wormhole, we use the symmetric
radial coordinate
\begin{equation}
R(r)=\sqrt{r^2+4M^2},
\end{equation}
with range $-\infty<r<\infty$, and impose isometry boundary conditions
at $r=0$.


\section{Reduction of the NOR formulations to spherical symmetry}
\label{Appendix:Reduction}

 
To our knowledge, little numerical work using the BSSN or NOR
formulations in spherical symmetry has been published, and therefore
we would like to point out a technical detail in the reduction to
spherical symmetry. $f_i$ can be defined as a true 3-vector by
introducing a flat auxiliary connection \cite{Meudonharmonic}, so that
\begin{equation}
\label{fa}
f_a=\gamma^{bc}\hat\nabla_c\gamma_{ab}
-{\rho\over 2}\gamma^{bc}\hat\nabla_a\gamma_{bc}.
\end{equation}
Here $\hat\nabla$ is defined so that its connection coefficients
vanish in preferred Cartesian coordinates $x^i$. In an evolution with
a single Cartesian coordinate patch these are the only coordinates,
and $\hat\nabla_a$ reduces to the partial derivative,
but if multiple Cartesian patches are used, one of them is preferred.

In spherical symmetry, $\hat\nabla_a$ is defined as the covariant
derivative compatible with the flat metric $ds^2=dr^2+r^2\,d\Omega^2$.
The result is
\begin{equation}
\label{B3}
f_r={\gamma_{rr}'\over \gamma_{rr}}+{2\over r}\left({\gamma_{rr}\over
  \gamma_T}-1\right)
-{\rho\over 2}\left({\gamma_{rr}'\over\gamma_{rr}}+2{\gamma_T'\over \gamma_T}\right)
\end{equation}
where $\gamma_T\equiv R^2/r^2$, we have written $\gamma_{rr}$
explicitly instead of just $\gamma$ as elsewhere in this paper, and
$f_\theta=f_\varphi=0$ because of spherical symmetry. Local
regularity at the origin $r=0$ requires $\gamma_{rr}-\gamma_T=O(r^2)$ and
$f_r=O(r)$, and so $f_a$ is a regular vector field. [Naively calculating
$f_r$ using partial derivatives in $(r,\theta,\varphi)$ gives a
different, singular result.]


\section{Spherical Einstein-scalar code}
\label{section:collapsecode}


We choose the collapsing matter to be a massless,
minimally coupled scalar field $\psi$ so the equations of motion
become
\begin{eqnarray}
{R_{ab}} &=& {\nabla _a} \psi {\nabla _b} \psi,
\\
{\nabla ^a}{\nabla _a} \psi &=& 0.
\end{eqnarray}
We will find it helpful to introduce the quantities $A$, $P$, $\Pi$, $s$ and
$w$ by
\begin{eqnarray}
A = {K_{rr}} -{\textstyle {\frac 1 3}} K,
\\
P={\psi '},
\\
{\Pi} = {n^a}{\nabla _a} \psi,
\\
s={\alpha '},
\\
w = {R'},
\end{eqnarray}
and also to define the quantities $\tilde \alpha$ and $\tilde s$ by
\begin{eqnarray}
{\tilde \alpha} = \ln \alpha,
\\
{\tilde s} = s/\alpha.
\end{eqnarray}

The spatial metric evolves by
\begin{equation}
{\partial _t} {\gamma_{ij}} = - 2 \alpha {K_{ij}} + {{\cal L}_\beta}
{\gamma _{ij}}.
\label{dtgamma}
\end{equation}
Then noting that ${\gamma _{rr}}=1$ we find that the two components 
of (\ref{dtgamma})
become
\begin{eqnarray}
0 = - \alpha \left ( A+{\textstyle {\frac 1 3}} K \right ) + {\beta '},
\\
{\dot R} = \beta w + \alpha R \left ({\textstyle {\frac 1 2}} A -
{\textstyle {\frac 1 3}} K \right ).
\label{dtR}
\end{eqnarray}
The first of these equations can immediately be integrated to yield
\begin{equation}
\beta = \int \alpha \left ( A + {\textstyle {\frac 1 3}} K \right ) d r.
\label{betaint}
\end{equation}
Differentiating the definition of $P$ with respect to time we obtain
\begin{equation}
{\dot P} = \beta {P '} + s \Pi + \alpha \left [ {\Pi '} + P \left (
{\textstyle {\frac 1 3}} K + A \right ) \right ],
\label{dtP}
\end{equation}
while the wave equation for $\psi$ yields
\begin{equation}
{\dot \Pi} = \beta {\Pi '} + s P + \alpha \left ( K \Pi + {P '} + 
{\frac {2 w P} R} \right ).
\label{dtPi}
\end{equation}
The momentum and Hamiltonian constraints yield respectively
\begin{eqnarray}
{A '} &= &{\frac {- 3 w A} R} - \Pi P + {\textstyle {\frac 2 3}} {K '},
\label{conm}
\\
{w'} &= &{\frac {1 - {w^2}} {2R}} - {\frac R 4} \left ( {\Pi ^2} + {P^2} + 
{\textstyle {\frac 3 2}} {A^2} - {\textstyle {\frac 2 3}}{K^2}\right ).
\end{eqnarray}
The momentum constraint enables us to differentiate 
(\ref{dtR}) with respect to $r$
yielding
\begin{equation}
{\dot w} = \beta {w '} + s R \left ( {\textstyle {\frac 1 2}} A - 
{\textstyle {\frac 1 3}} K \right ) - {\textstyle {\frac 1 2}}
\alpha R P \Pi .
\label{dtw}
\end{equation}
The BMn slicing condition is
\begin{equation}
{\dot {\tilde \alpha}} = \beta {\tilde s} - 2 K,
\label{dtalphat}
\end{equation}
which when differentiated with respect to $r$ yields
\begin{equation}
{\dot {\tilde s}} = \beta {{\tilde s} '} + \alpha {\tilde s} \left ( 
A + {\textstyle {\frac 1 3}} K 
\right ) - 2 {K '}.
\label{dtst}
\end{equation}

The extrinsic curvature evolves by
\begin{eqnarray}
\nonumber
{{\cal L}_t} {{K^a}_b} = {{\cal L}_\beta} {{K^a}_b} - {D^a}{D_b}\alpha
\\
+ \alpha ({^{(3)}}{{R^a}_b} + K {{K^a}_b} - {D^a}\psi {D_b}\psi )  ,
\label{dtKab}
\end{eqnarray}
where $D_a$ and ${^{(3)}}{{R^a}_b} $ are respectively the derivative
operator and Ricci tensor associated with the spatial metric 
$\gamma _{ab}$.  Eq.~ (\ref{dtKab}) yields the evolution equations
\begin{eqnarray}
{\dot K} = \beta {K '} - {s '} - {\frac {2 w s} R} + \alpha \left (
{\textstyle {\frac 1 3}} {K^2} + {\textstyle {\frac 3 2}} {A^2} +
{\Pi ^2} \right ) 
\label{dtK}
\\
\nonumber
{\dot A} = \beta {A '} + {\textstyle {\frac 2 3}}\left ( {\frac {w s} R}
- {s '} \right )
\\
+ \alpha \left ( {\textstyle {\frac 1 6}} {\Pi ^2} - 
{\textstyle {\frac 1 2}} {P^2} + {\textstyle {\frac 1 4}} {A^2} 
- {\textstyle {\frac 1 9}} {K^2} + K A + {\frac {{w^2} -1} {R^2}} \right ).
\label{dtA}
\end{eqnarray}
We evolve this system in one of two different ways depending on whether 
the numerical inner boundary is a regular centre or an excision
boundary.

If the inner boundary is a regular centre then the variables to be
evolved are ($R,P,\Pi ,w,{\tilde \alpha},{\tilde s},K$) which are 
evolved using 
Eqs.~(\ref{dtR},\ref{dtP},\ref{dtPi},\ref{dtw},\ref{dtalphat},\ref{dtst},
\ref{dtK})
respectively.  
In the evolution equations, the quantities $\alpha$, $s$ and $s'$ are
regarded as derived from the evolved variables 
$\tilde\alpha$ and $\tilde s$.
The shift $\beta$ is determined from (\ref{betaint}) 
and the condition that $\beta$ vanish at the centre. Similarly $A$ is 
computed by integrating (\ref{conm}) outward from the centre and
using the fact that $A$ vanishes at the centre.  Boundary conditions at
the centre simply follow from smoothness of the metric which requires
that $R,P$ and $\tilde s$ vanish there, that the spatial derivatives of
$\Pi$, ${\tilde \alpha}$ and $K$ vanish, and that $w=1$.     

If the numerical inner boundary is an excision boundary, then the
variable $A$ is not computed by the momentum constraint but is instead
evolved using (\ref{dtA}).  The integration constant for
\ref{betaint}) is determined by having the time derivative
of the area radius $R$ vanish at the apparent horizon.  At the
excision boundary, all modes are outgoing, except for a single
incoming gauge mode $s-\sqrt{2\alpha}K$. As a boundary condition, its
value is prescribed as the value it had at the previous time step.
All other modes are evolved at the excision boundary point; however
spatial derivatives at that point are computed using one sided
differences rather than the centred differences used at the other
points \cite{Pretorius2}.

For initial data we choose a moment of time symmetry so that $K$ and
$\Pi$ vanish.  The initial value of $P$ is given by
\begin{equation}
P= a {r^3}\exp(-{r^2}/{\sigma ^2})  
\end{equation}
where $a$ and $\sigma$ are constants, 
and the initial value of $R$ is solved for by using the Hamiltonian 
constraint. The lapse $\alpha$ is given an initial value of 1.
The grid points are equally spaced in $r$.  
Spatial derivatives are evaluated using standard second 
order centred differences and time evolution is done by the iterative
Crank-Nicholson method.  A marginally outer trapped surface 
occurs where the derivative of $R$ along the outgoing null direction
vanishes.  In terms of the variables used here, that corresponds to
the vanishing of $w+R({\frac 1 2} A - {\frac 1 3} K)$.    
If the apparent horizon (the outermost marginally outer trapped surface)
occurs at gridpoint $i$ (and if we choose to excise), 
then we place the excision boundary at gridpoint
$9i/10$. The constants determining
the initial value of $P$ were chosen to be $a=0.05$ and ${\sigma ^2}=8$.  
The number of spatial gridpoints was chosen to be 6401 and the 
initial range of $r$ was chosen to be $0 \le r \le 200$. 



\end{document}